\documentclass{svproc}
\usepackage{color}
\usepackage{comment}

\mathchardef\mhyphen="2D

\usepackage[colorinlistoftodos]{todonotes}
\usepackage{lipsum}
\usepackage{amsmath}
\usepackage{amssymb}
\usepackage[ruled,vlined,linesnumbered,lined,boxed]{algorithm2e}

\newcommand{\TOPK}{{\sf Weighted-Top-k-Overlapping DCS}}

\newcommand{\IDS}{{\sf IWDS}}

\usepackage{url}

\begin{document}
\mainmatter              
\title{Top-k Connected Overlapping Densest Subgraphs in Dual Networks}

%
\titlerunning{Top-k overlapping densest connected subgraphs in Dual Networks}  
%
\author{Riccardo Dondi\inst{1} \and Pietro Hiram Guzzi\inst{2}
 \and
Mohammad Mehdi Hosseinzadeh\inst{1}}
\authorrunning{Riccardo Dondi et al.} 
%
\tocauthor{Riccardo Dondi, Pietro Hiram Guzzi and Mohammad Mehdi Hosseinzadeh}
\institute{Universit\`a degli Studi di Bergamo, Bergamo, Italy,\\
\email{riccardo.dondi@unibg.it}; \email{m.hosseinzadeh@unibg.it},\\
Magna Graecia University, Catanzaro, Italy,\\
\email{hguzzi@unicz.it}
}

\maketitle              

\begin{abstract}
Networks are largely used for modelling and analysing data and relations among them.  
Recently, it has been shown that the use of a single network may not 
be the optimal choice, since a single network may misses some aspects. Consequently, it has been proposed to use a pair of networks to better model all the aspects, and the main approach is referred to as dual networks (DNs). DNs are two related graphs (one weighted, the other unweighted) that share the same set of vertices and two different edge sets. 
In DNs is often interesting to extract common subgraphs among the two networks that are maximally dense in the conceptual network and connected in the physical one. The simplest instance of this problem is finding a common densest connected subgraph (DCS), while we here focus on the detection of the Top-k Densest Connected subgraphs, i.e. a set k subgraphs having the largest density in the conceptual network which are also connected in the physical network. 
We formalise the problem and then we propose a heuristic to find a solution,
since the problem is computationally hard. A set of experiments on synthetic and real networks is also presented to support our approach. 
\end{abstract}

\section{Introduction}

Networks are largely used to represent and analyse data and relations among them in many fields \cite{Cannataro2010}. For instance, in biology and medicine networks are used to model relationships among macromolecules in a cell (e.g. nucleic acids, proteins and genes). Similarly, in social network analysis, graphs are used to model associations among users and the analysis may reveal association patterns or communities of similar users \cite{sapountzi2018social}.
Classically, a single network has been used to model data and to extract relevant knowledge by looking at topological parameters, i.e. community-related structures \cite{cannataro2010impreco} such as groups of related genes or the presence of related users \cite{liu2018d}. More recently, it has been shown that the use of a single network may not be able to capture efficiently all the relationships among modelled entities, therefore some complex models have been introduced such as heterogenous networks \cite{hetnetaligner} or dual networks \cite{Wu:2016tx}. In particular  dual networks are two  related graphs sharing the same vertex set and two different edge sets. 
One network has unweighted edges, and it is called \textit{physical graph}. The second one has weighted edges and it is called \textit{conceptual graph}. Dual networks have been used in other works to model interactions among genetic variants \cite{Phillips:2008dm}. 





\begin{figure}
    \centering
    \includegraphics[width=2.8 in]{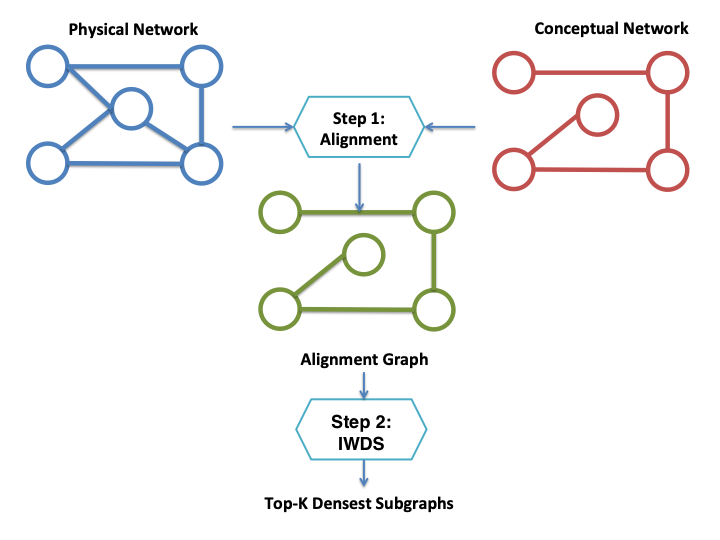}
    \caption{Workflow of the proposed approach.  In the first step the input conceptual and physical networks are merged together using a network alignment approach; then \TOPK{} are extracted from the alignment graph. Each extracted subgraph induces a connected subgraph in the physical network and one of the top-k overlapping weighted densest subgraph in the conceptual one.}
    \label{fig:dcsasgraph}
\end{figure}

In dual networks, finding a Densest Connected Subgraph
(DCS) may reveal hidden and relevant knowledge that cannot captured by using a single graph. Formally, given two input graphs $G_{p}(V,E_{p})$ (undirected and edge-weighted), and $G_{c}(V,E_{c})$ (undirected and unweighted), the problem consists in finding a subset of nodes $ I_{s}$ that induces a densest community in $G_{c}$ and a connected subgraph in $G_{p}$.  As proved in \cite{Wu:2016tx} the DCS problem is NP-hard in its general formulation 
\cite{Karp:2009ko}, therefore there is the need for introducing novel heuristics and approaches able to solve it. 

Recently, researchers focused on the search of a set of densest subgraphs that may better capture the presence of communities in real networks \cite{DBLP:conf/wsdm/BalalauBCGS15,DBLP:journals/datamine/GalbrunGT16}.
Many classical problems looks for a single dense 
subgraph, while in \cite{DBLP:journals/datamine/GalbrunGT16,TOPK2019} the problem has been extended to a set of $k$,
with $k\geq 1$, densest subgraphs. Such overlapping subgraphs may represent relevant communities sharing some nodes, such as hubs. We here explore the problem of finding top-k weighted overlapping densest connected subgraphs in dual networks.
While finding such graphs in a single network has been resolved by ad-hoc heuristics \cite{lee2010survey}, the variant of the problem in dual networks is still a challenging problem. Therefore, we model the problem as a variant of the local network alignment problem and we propose a novel algorithm to solve it. The approach is based on a two step strategy: first a single alignment graph is built from the dual networks \cite{PietroHiramGuzzi:2017bn,hetnetaligner}, then we mine such graph with an ad-hoc heuristic. 


The extracted subgraphs on the alignment graph induce dense subgraphs on the conceptual networks and  connected subgraphs in the physical one, therefore they are solutions of the initial problem. 
Figure \ref{fig:dcsasgraph} depicts the workflow of our approach.

With respect to the existing approaches for 
the DCS problem, our approach is conceptually different and it enables more flexibility.  For instance, Wu et al \cite{Wu:2016tx} do not consider overlapping subgraphs and their approach is limited to the exact correspondence of nodes between networks, while our approach may also find other kind of communities (e.g. by using a different algorithm for mining alignment graph).
On the other hand, with respect to other approaches
for finding densest subgraphs in a network
\cite{DBLP:conf/wsdm/BalalauBCGS15,DBLP:journals/datamine/GalbrunGT16,TOPK2019,guzzi2010mu}, 
we consider weighted networks, 
an extension that can be useful in many contexts,
in particular for biological networks.

We provide an implementation of our heuristic and we show the effectiveness of our approach on synthetic
datasets and through three case studies on social networks data, on biological networks and on a co-authorship networks. The experimental 
results confirm the effectiveness of our approach.

The paper is structured as follows: Section \ref{sec:related} discusses main related works
and Section \ref{sec:def} gives definitions related
to our problem and formally introduces the
problem we are interested into.
Section \ref{sec:algo} presents our heuristic; Section \ref{sec:experiments} discusses the case studies; finally Section \ref{sec:conclusion} concludes the paper.


\section{Related Work}
\label{sec:related}


Many real life contexts cannot be efficiently modelled using a single network without losses of information, therefore the use of dual networks is growing \cite{Wu:2016tx}. 
The problem of finding communities in a 
network or a dual network is based on the specific model
of dense or cohesive graph considered.
Several models of cohesive subgraph have been considered in literature and applied in different  contexts.
One of the first definition of cohesive subgraph is a fully connected subgraph, i.e. a clique. However, the determination of a maximal clique, also referred to as the Maximum Clique problem, is NP-Hard \cite{hastad1996clique}, and it is difficult to approximate \cite{DBLP:conf/stoc/Zuckerman06}. Moreover, real networks sometimes miss some edges, therefore the clique model is often too strict and may fail to find some important information.
Consequently, many alternative definitions of cohesive subgraphs that are not fully interconnected have been introduced, including 
densest subgraph \cite{DBLP:journals/algorithms/Komusiewicz16}.


A \textit{densest subgraph} is a subgraph with maximum density (ratio between number of edges 
and number of nodes of the subgraph)
and the  Densest-Subgraph problem asks for a subgraph of maximum density in a given graph. The problem can be solved in polynomial time 
\cite{goldberg1984finding,DBLP:journals/algorithmica/KawaseM18} 
and approximated within factor $\frac{1}{2}$
\cite{asahiro2000greedily,DBLP:conf/approx/Charikar00}. 
Notice that the Densest-Subgraph problem can be extended
also to edge-weighted networks.

Recently, Wu et al. \cite{Wu:2016tx} , proposed an algorithm for finding a densest connected subgraph in a dual network. The approach is based on a two-step strategy. In the first step the algorithm prunes the dual network without eliminating the optimal solution. In the second step two greedy approaches are developed to build a search strategy for finding a densest connected subgraph. Briefly, the first approach finds a densest subgraph in the conceptual network first, and then it is refined to guarantee that it is connected in the physical network. The second approach maintain the subgraph connected in the physical network while deleting low-degree nodes in the conceptual network. 



While the literature of network mining has mainly focused on
the problem of finding a single subgraph, 
recently the interest in finding more than
a subgraph has emerged~\cite{DBLP:conf/wsdm/BalalauBCGS15,DBLP:journals/datamine/GalbrunGT16,TOPK2019,hosseinzadeh2020dense}. 
The proposed approaches usually allows 
overlapping between the computed dense subgraphs.
Indeed, there can be nodes that are shared between
interesting dense subgraphs, for example hubs.
The proposed approaches differ on the way overlapping
is handled. 
The problem defined in~\cite{DBLP:conf/wsdm/BalalauBCGS15} controls the
overlap by limiting the Jaccard coefficient 
between each pair of subgraphs of the solution.
The Top-k-Overlapping problem introduced in
~\cite{DBLP:journals/datamine/GalbrunGT16}
includes a distance function into the objective function.
In this paper, we follow this last approach and we extend it 
to weighted networks. 

\section{Definitions}
\label{sec:def}

In this section we define the main concepts 
related to our problem starting with the definition of dual network.

\begin{definition} Dual Network.\\
A dual network DN $G(V,E_\phi,E_{con})$ is a pair of networks: a conceptual weighted network $G_{con}(V,E_{con})$ and a physical unweighted one $G_\phi(V,E_\phi)$.
\end{definition}


Now, we introduce the definition of 
density.

\begin{definition}Density.\\
Given a weighted graph $G(V,E,weight)$, let $v \in V$ 
be a node of $G$, and 
\[vol(v)=\sum_{w:(v,w)\in E}weight(v,w)\] be the sum of the weights of the edges. The density of weighted graph is defined as $\rho(G)=\frac{\sum_{v \in V}vol(v)}{|V|}$.
\end{definition}

Given a graph (weighted or unweighted) $G$
with a set $V$ of nodes and a subset $Z \subseteq V$,
we denote by $G[Z]$
the subgraph of $G$ induced by $Z$.
Given $E' \subseteq E$, we denote by
$weight(E')$ the sum of weights of edges in $E'$.

Given a dual network we may consider the subgraphs $G_{p}[I]$ and $G_{c}[I]$ induced in the two networks by the same node set $I \subseteq V$.
A densest common subgraph $DCS$ is a subset of nodes $I$ such that the density of the induced conceptual network is maximised and the induced physical network is connected, formally defined
in the following.

\begin{definition}Densest Common Subgraph.\\ 
Given a dual network $G(V,E_c,E_p)$, a densest connected subgraph in $G(V,E_c,E_p)$ is a subset of nodes $I \subseteq V$ such that $G_p[I]$ is connected and the density of $G_c[I]$ is maximised.
\end{definition}

In this paper, we are interested in finding 
$k \geq 1$  densest connected subgraphs.
However, to avoid taking the same copy
of a subgraph or subgraphs
that are very similar, we consider
the following distance functions
introduced in \cite{DBLP:journals/datamine/GalbrunGT16}.

\begin{definition}
Let $G(V,E_c,E_p)$ be a dual network and let $G[A]$, $G[B]$, 
with $A, B \subseteq V$, be two induced subgraphs of $G$. 
The distance between $G[A]$ and $G[B]$, denoted by
$d: 2^{V} \times 2^{V} \rightarrow \mathbb{R_{+}}$ 
has value equal $2-\frac{|A \cap B|^2}{|A||B|}$  if  $A \neq B$,
else is equal to $0$.
%

\end{definition}

Now, we are able to introduce the problem we are interested into.

\begin{problem}\TOPK{} \\
\noindent
\textbf{Input:} A dual network $G(V,E_c,E_p)$, a parameter $\lambda > 0$. \\
\textbf{Output:} a set $\mathcal{X} = \{ G[X_1], \dots , G[X_k] \}$
of $k$ connected subgraphs of $G$, 
with $k \geq 1$, such that the
following objective function is maximised:
\[
\sum_{i=1}^{k} \rho(G_c[X_i])+ \lambda \sum_{i=1}^{k-1} \sum_{j=i+1}^k d(G[X_i],G[X_j]) 
\]
\end{problem}

\TOPK{}, for $k \geq 3$, is NP-Hard, as 
it is NP-Hard already on an unweighted
graphs \cite{TOPK2019}.
Notice that for $k=1$, then \TOPK{} is
exactly the problem of finding a single weighted densest
connected subgraph, hence it can be solved
in polynomial time.

\subsection{Greedy Algorithms for DCS}
\label{subsect:Greedy}

One of the ingredient of our method is a variant of a greedy algorithm for DCS.
This latter algorithm,
denoted by \textsf{Greedy},
is an approximation algorithm for 
computing a connected densest subgraph of a given graph. 
Given a weighted graph $G$, 
\textsf{Greedy} \cite{asahiro2000greedily,DBLP:conf/approx/Charikar00}
iteratively removes from $G$ a vertex $v$ having lowest $vol(v)$ and stops when all the vertices
of the graph have been removed.
At each iteration $i$, with $1 \leq i \leq |V|$, 
\textsf{Greedy} computes a subgraph $G_i$
and returns a densest of subgraphs
$G_1, \dots , G_{|V|}$.
The algorithm has a time complexity 
$O(|E| + |V| \log |V|)$ on a weighted graph
and achieves an approximation factor of
 $\frac{1}{2}$ \cite{asahiro2000greedily,DBLP:conf/approx/Charikar00}. 
 
We introduce here a variant of the \textsf{Greedy} algorithm,  called \textsf{V-Greedy}.
Given an input weighted graph $G$,
similar to \textsf{Greedy}, \textsf{V-Greedy}
iteratively removes a vertex $v$
having lowest $vol(v)$
and at each iteration $i$, with $1 \leq i \leq |V|$, computes a subgraph $G_i$,
with $1 \leq i \leq |V|$.
Then, among subgraphs $G_1, \dots , G_{|V|}$, 
\textsf{V-Greedy} returns a subgraph $G_i$
that maximises the value $\rho(G_i) + 2(\frac{\rho(G_i)}{|V_i|})$.

Essentially, we add a correction factor
$2(\frac{\rho(G_i)}{|V_i|})$ to the density
to avoid that the returned
subgraph consists of almost disjoint dense subgraphs.
For example, consider 
a graph with two equal size cliques 
$K_1$ and $K_2$ having the same (large) weighted density and
a single edge of large weight connecting them. 
Then
the union of $K_1$ and $K_2$ is denser than both $K_1$
and $K_2$, hence \textsf{Greedy} returns
the union of $K_1$ and $K_2$ and this may prevent
us to find  $K_1$, $K_2$ as a solution of \TOPK{}.
\textsf{V-Greedy}, when the density
of $K_1$ and $K_2$ is close enough to
the density of their union, will return instead one of $K_1$, $K_2$.

\section{The Proposed Algorithm} 
\label{sec:algo}

In this section we present the proposed heuristic approach
for \TOPK{} in dual networks.
The approach
is based on two main steps: (i) First, the input networks are integrated into a single  weighted alignment graph preserving the connectivity properties of the physical networks; (ii) Second, the obtained alignment graph is mined by using an ad-hoc heuristic for \TOPK{}. 

\subsection{Building of the Alignment Graph}
\label{sec:building}

In the first step the algorithm receives as input a weighted graph $G_1(V,E_1)$, and an unweighted graph  $G_2(V,E_2)$,  
an initial set of node pairs between input networks  (\textit{seed nodes}) representing corresponding nodes,  a distance threshold $\delta$ that represents the maximum threshold of distance that two nodes may have in the physical network (the parameter is optional and it is used to prune the possible solutions). 

Starting from these input data, the algorithm starts by building the nodes of the alignment graph. Each node of the input graph represents a pair of corresponding nodes of the input ones following the \textit{seed nodes}. Then it adds the edges among nodes. The algorithm adds an edge connecting two nodes whenever the corresponding nodes are connected in both input networks. If the nodes are adjacent in both networks, then the edge connecting them will have  the same weight of the corresponding nodes in the conceptual network.
Conversely, if the nodes are adjacent in the physical network and they have a distance lower than $\delta$ (an user defined threshold of distance), then the weight of the edge will be the average of the weights of the path linking the considered nodes.

\subsection{ A Heuristic for \TOPK{}}
\label{sec:heuristic}

In this section, we present our heuristic for \TOPK{},
called {\sf Iterative Weighted Dense Subgraphs} (\IDS{}).
\IDS{} is based on \textsf{V-Greedy} 
(see Section \ref{subsect:Greedy}), applied iteratively
to compute $k$ subgraphs.

The heuristic starts with 
$\mathcal{X}= \emptyset$ and 
consists of $k$ iterations.
At each iteration $i$, $1 \leq i \leq k$,
given a set
$\mathcal{X}= \{G[X_1],...,G[X_{i-1}]\}$ of subgraphs, 
\IDS{} 
computes a subgraph $G[X_i]$ and adds it to $\mathcal{X}$.

The first iteration of \IDS{} applies the \textsf{V-Greedy} 
algorithm on the dual network $G$ 
and computes $G[X_1]$. In iteration $i$, with $2 \leq i \leq k$, \IDS{} applies one of the two following cases, depending on a parameter $f$,
$0 < f \leq 1$,
and on the size of the set $C_{i-1} = \bigcup_{j=1}^{i-1} X_j$
(the set of nodes covered by the subgraphs
in $\mathcal{X}$). 

\paragraph{Case 1.} If $|C_{i-1}| \leq f |V|$
(at most  $f |V|$ nodes of $G$ are covered),
\IDS{} applies the \textsf{V-Greedy} 
algorithm on a graph obtained
by retaining  $\alpha$ nodes ($\alpha$ is a parameter) 
of $C_{i-1}$ having highest weighted degree in $G$
and removing the other nodes of $C_{i-1}$.
$G[X_i]$ is a weighted connected dense subgraph,
distinct from those in $\mathcal{X}$,
in the resulting graph. 

\paragraph{Case 2.} If $|C_{i-1}| > f |V|$
(more than $f |V|$ nodes of $G$ are covered),
\IDS{} applies the \textsf{V-Greedy} 
algorithm on a graph obtained
by removing $(1-\alpha)$ nodes (recall that $\alpha$ is a  parameter of \IDS{}) of $C_{i-1}$ having  
lowest weighted degree in $G$.
\IDS{} computes $G[X_i]$ as a weighted 
connected dense subgraph, distinct from those in $\mathcal{X}$,
in the resulting graph.


\paragraph{Complexity Evaluation.}



We denote by $n$ (by $m$, respectively) the number of nodes (of edges, respectively) of the dual network. 
The first step requires the analysis of both graphs and the building of the novel alignment graph.  
The construction of the alignment graph requires $\mathcal{O}(n^2)(calculation-edge-weights)$ time.
The calculation of edge weights requires the calculation of the shortest paths among all the node pairs in the physical graph using the Chan implementation\cite{chanshortestpaths}, therefore it requires $\mathcal{O}(n m_p)$ ($m_p$ is the number of edges of the physical graph). 


As for Step 2, \IDS{} makes $k$ iterations.
Each iteration
applies \textsf{V-Greedy} on $G$
and requires $O(m n \log n)$ time \cite{charikar2000greedy}.
Iteration $i$, with $2 \leq i \leq k$,
computes the set of covered nodes to retain
or remove, 
which requires $O(n \log n)$ time 
by sorting
the nodes in $C_{j-1}$ based on their weighted degree.
Thus the overall time complexity of \IDS{} 
is $O(k m n \log n)$. 

\section{Experiments}
\label{sec:experiments}

In this section, we provide an experimental evaluation of our heuristic \IDS{} on synthetic and real networks\footnote{The source code and data used in our experiments are available at \url{https://github.com/mehdihosseinzadeh/-k-overlapping-densest-connected-subgraphs}}. The design of a strong evaluation scheme for our algorithm is not simple since we have to face two main problems: (1)
    Existing methods for 
    computing the top-k overlapping subgraphs are defined for unweighted graphs
    and cannot be used on dual networks; 
(2)    Existing network alignment algorithms do not aim to extract densest subgraphs.

Consequently, we cannot easily compare our approach with the existing state of the art methods, and we design an ad-hoc procedure for the evaluation based on the following steps.  First, we demonstrate on synthetic networks that our approach can find densest known subgraphs. In this way, we show that \IDS{} can correctly recover top-k weighted densest subgraphs. Then we applied our method to some real-world dual networks. 

The alignment algorithm is implemented in  Python 3.7 using the NetworkX package for managing networks \cite{hagberg2008exploring}. We implemented \IDS{} in MATLAB R2020a and
we perform the experiments on MacBook-Pro (OS version 10.15.3) with processor 2.9 GHzIntel Core i5 and 8GB 2133 MHz LPDDR3 of RAM, Intel Iris Graphics 550 1536 MB.
We consider the value of parameter $\alpha$ equal to range from 0.05 to 0.25.

\subsection{Synthetic Networks}

In the first part of our experimental evaluation, we analyse the performance of \IDS{} to find planted ground-truth subgraphs on synthetic datasets.

\textbf{Datasets.} We generate a synthetic dataset consisting
of $k=5$ planted dense subgraphs (cliques
or graphs close to cliques). 
A planted dense subgraph contains five disjoint
cliques of $30$ nodes each, whose edges
have weights randomly generated
in the interval $[0.8,1]$. These cliques
are then connected to a background subgraph
of $100$ nodes. We consider three different
ways to generate the background subgraph:
Erd{\"o}s-Renyi with parameter $p=0.1$,
Erd{\"o}s-Renyi with parameter $p=0.2$
and Barabasi-Albert with parameter equal to $10$.
Weights of the background graphs are randomly generated
in interval $[0,0.5]$.
Then $50$ edges are randomly added (with 
weights randomly generated in interval $[0,0.5]$).

Based on this approach, we generate two different sets of synthetic networks, called  \emph{Synthetic1} and \emph{Synthetic2}. 
\emph{Synthetic1} is generated as described above,
while in \emph{Synthetic2} noise is added
to modify the synthetic networks.
In this latter case, noise is added by varying $5\%$ and $10\%$
of node relations. Pairs of nodes are chosen randomly:
if they 
belong to the same clique, 
the weight of the edge connecting them
is changed to a random value in the interval $[0,0.5]$;
else 
and edge connecting them is (possibly) added
(if not already in the network)
and weight is randomly assigned a value in
the interval $[0.8,1]$.

\textbf{Outcome.} Tables \ref{table:synth1} 
reports 
F1-score\footnote{Following \cite{yang2012community}, each ground-truth subgraph and each detected subgraph are compared based on their nodes, hence we compute the F1-scores by determining which ground-truth subgraph corresponds to which detected subgraph.}
as a measure of a test's accuracy to detect the ground-truth subgraph by taking into account precision and recall of the test to compute the score. 
The F1-score takes value between 0 and 1, 
where the value 1 is obtained when the detected subgraphs exactly correspond to the ground-truth subgraphs.

We apply an approach similar to \cite{yang2012community} to compare the sets of subgraphs detected by the \IDS{} to the planted ground-truth by using the $F1[t/d]$ measure
as the average F1-score of the best-matching ground-truth subgraph to each detected subgraph (\textit{truth to detected}) and $F1[d/t]$ measure as the average F1-score of the best-matching detected subgraph to each ground-truth subgraph (\textit{detected to truth}). Table \ref{table:synth1} shows that for the noiseless \emph{Synthetic1} \IDS{} is able to find the ground-truth subgraphs for all three different values of $\alpha$, averaged over 300 examples.

Table \ref{table:synth2} shows performance of \IDS{} on the \emph{Synthetic2} datasets using noise 0.05 and 0.10, averaged over 90 examples. For noise 0.05, \IDS{}  outputs solutions that are very close to the ground-truth subgraphs for all three different values of $\alpha$. With the increasing of noise to 0.10 the algorithm performance is slightly worse, but still remains close to optimal solutions.





\begin{table}[htb]
\centering 
\caption {Performance of \IDS{}:
(on the left) performance on \textit{synthetic1} for $k=5$, varying $\alpha$ from 0.05 to 0.25, averaged over 300 examples; 
(on the right) performance of \IDS{} on \textit{synthetic2} for $k=5$, varying $\alpha$ from 0.05 to 0.25, averaged over 90 examples.}
\begin{tabular}{|l|c|c|c|} 
\hline 
&$\alpha=0.05$&$\alpha=0.1$&$\alpha=0.25$\\ 
\hline
$F1[t/d]$&1.00&1.00&1.00\\
$F1[d/t]$&1.00&1.00&1.00\\

\hline 
\end{tabular}
\quad
\quad
\begin{tabular}{|l|l|c|c|c|} 
\hline 
Noise&&$\alpha=0.05$&$\alpha=0.1$&$\alpha=0.25$\\ 
\hline
0.05&$F1[t/d]$&0.99&1.00&1.00\\
&$F1[d/t]$&0.99&0.99&0.99\\
\hline
0.10&$F1[t/d]$&0.98&0.97&1.00\\
&$F1[d/t]$&0.96&0.96&0.97\\

\hline 
\end{tabular}
\label{table:synth1}
\label{table:synth2}
\end{table}

\subsection{Dual Networks}

We now evaluate \IDS{} on three following real-world dual network datasets:

\textbf{Datasets.} 
\textit{G-graphA}. The G-graphA dataset is derived from the GoWalla  social network where users share their locations (expressed as GPS coordinates) by checking-in into the web-site \cite{cho2011friendship}. Each node represents a user and each edge link two friends into the network.  We obtained the physical network by considering friendship relation on the social network. We calculated the conceptual network by considering the distance among users. Then we run the first step of our algorithm and we obtained the alignment graph \textit{G-graphA},
containing 2241339 interactions and 9878 nodes (we set $\delta$=4). In this case a DCS represents set of friends that share some near locations of check-ins, i.e. communities of related users.


\textit{DBLP-graphA}. The \textit{DBLP-graphA} dataset is a computer science bibliography represents interaction between the authors. Nodes represent authors and edges represent connections between two authors if they publish at least one paper together. Each edge in the physical network connect two authors that co-authored at least a paper. Edges in the conceptual network represent the similarity of research interest of the authors calculated on the basis of all their publications. After running the first step of the algorithm (using $\delta$=4), we obtained an alignment graph  \textit{DBLP-graphA} dataset containing 553699 interactions and 18954 nodes.  In this case
a DCS represents a set of co-authors that share some strong common research interests and the use of DNs is mandatory, since physical network shows only co-authors that may not have many common interests and the conceptual network represents authors with common interest that may not be co-authors.

\textit{HS-graphA}. The \textit{HS-graphA} dataset is a biological data 
and is taken from the STRING database \cite{stringdatabase}. Each node represents a protein, and each edge takes into account the reliability of the interaction. We use two networks for modelling the database: a conceptual network representing such reliability value; and a physical network storing the binary interactions. The \textit{HS-graphA} dataset contains 5879727 interactions and 19354 nodes (we set $\delta$=4).

\begin{table}[]
    \caption{Properties of the Alignment Graphs obtained for each dataset.}
    \centering
    \begin{tabular}{|c|c|c|c|c|}\hline
        Graph & Representation	&	Nodes	&	Edges	&	density	\\ \hline
DBLP-graphA & co-authorship  & 18954	&	553699	&	0.0026	\\ \hline
G-graphA & Socialì	& 9878	&	2241339	&	0.0448\\ \hline
HS-graphA & Protein Interactions	&	19354	&	5879727	&	0.0313	\\ \hline
    \end{tabular}
    \label{tab:graphpropertiesn}
\end{table}

\textbf{Outcome.} For these large size datasets, we set the value of $k$ to $20$,
following the approach in \cite{DBLP:journals/datamine/GalbrunGT16}.
Table \ref{table:real1} reports the time, total density and distance returned by \IDS{} for three different values of $\alpha$. As shown in Table \ref{table:real1}, by increasing the value of $\alpha$ from $0.05$ to $0.25$, \IDS{} (except of one case, \textit{HS-graphA} with $\alpha=0.1$) returns
solutions that are denser, but with lower distance, hence these solutions consist of
subgraphs with more overlapping. 
Table \ref{table:real1} shows also how the running time of \IDS{} is influenced by the size of the network and by the value of $\alpha$. 
The running time is affected
not only by the number of nodes of the network,
but also by their density. DBLP-graphA and
HS-graph-A have almost the same number of nodes,
but HS-graph-A is much more denser than DBLP-graphA. \IDS{}
for the former network is considerably slower than for DBLP-graphA
($1.986$ slower for $\alpha =0.05$,
$6.218$ slower for $\alpha =0.25$).
The running time of \IDS{} 
increases as $\alpha$ increases.
This is due to the fact that, by increasing the
value of $\alpha$, less nodes are removed
by Case 1 and Case 2 of \IDS{}, hence
in iterations of \IDS{} \textsf{V-Greedy} is applied to larger subgraphs.

\begin{table}[htb]
\centering 
\caption {Performance of \IDS{} on real-world network for $k=20$, varying $\alpha$ from 0.05 to 0.25. For each network, we report the
			size of the network (number of vertices $|V|$ and 
			edges $|E|$), the running time in minutes, the density and the distance.}
\begin{tabular}{|l|c|c|l|c|c|c|} 
\hline 
set&$|V|$&$|E|$&&$\alpha=0.05$&$\alpha=0.1$&$\alpha=0.25$\\ 
\hline
&&&Time&89.84&98.72&184.87\\ 
G-graphA&9878&2241339&Density&2863.99&4000.73&6345.67\\ 
&&&Distance&275.82&257.84&220.16\\
\hline 
&&&Time&105.69&125.71&165.25\\ 
DBLP-graphA&18954&553699&Density&39.61&52.39&74.12\\ 
&&&Distance&307.72&231.25&213.04\\

\hline
&&&Time&209.88&749.06&1027.58\\ 
HS-graphA&19354&5879727&Density&1326.07&1153.68&1799.22\\ 
&&&Distance&226.40&212.34&205.55\\
  	     
\hline 
\end{tabular}
\label{table:real1}
\end{table}

\section{Conclusion}
\label{sec:conclusion}

DNs are used to model two kinds of relationships among objects in the same scenario. A DN is a pair of networks that have the same set of nodes. One network has unweighted edges (physical network), while the second one has weighted edges (conceptual network). In this contribution, we introduced
an approach that first integrates a physical
and a conceptual network into an alignment 
graph.
Then, we applied the \TOPK{} problem
to the alignment graph to find
$k$ dense connected subgraphs. These subgraphs represent some subsets of nodes that are strongly related in the conceptual network and that are connected in the physical one. We
presented a heuristic, called \IDS{}, for \TOPK{} and an
experimental evaluation of \IDS{}. We first presented as a proof-of-concept the ability of our algorithm to retrieve known densest subgraphs in synthetic networks. Then we tested the approach on some real networks  which demonstrate the effectiveness of our approach. Future work will regard a possible high performance implementation of the approach and the application of the \IDS{} algorithm to some other scenarios (e.g. financial or marketing datasets).



\bibliographystyle{splncs03}
\bibliography{Biblio.bib}




\end{document}